\documentclass[12pt]{article}

\usepackage{latexsym,amsmath,amssymb,fancybox}

\usepackage[english]{babel}

\usepackage{color,epsfig}
\usepackage{graphicx}

\definecolor{navy}{rgb}{0.0,0.0,0.4}

\definecolor{rd}{rgb}{1,0,0}
\definecolor{or}{rgb}{1,.33,0}
\definecolor{pi}{rgb}{.66,.33,.33}
\definecolor{gn}{rgb}{0,.50,0}
\definecolor{be}{rgb}{0,0,.66}
\definecolor{ru}{rgb}{.66,0,.33}
\definecolor{vi}{rgb}{.33,0,.66}
\definecolor{gy}{rgb}{0,.33,.66}
\definecolor{ye}{rgb}{.66,.33,0}
\definecolor{bk}{rgb}{0,0,0}

\def\thf{\baselineskip=\normalbaselineskip\multiply\baselineskip
by 7\divide\baselineskip by 6}

\def\fff{\baselineskip=\normalbaselineskip}

\def\spose#1{\hbox to 0pt{#1\hss}}
\def\lta{\mathrel{\spose{\lower 3pt\hbox
{$\mathchar"218$}}\raise 2.0pt\hbox{$\mathchar"13C$}}}  \def\gta{\mathrel
{\spose{\lower 3pt\hbox{$\mathchar"218$}}\raise 2.0pt\hbox{$\mathchar"13E$}}}

\def\sqr#1#2{{\vcenter{\hrule height.4pt\hbox{\vrule width.8pt height#2pt
\kern#1pt\vrule width.8pt}\hrule height.4pt}}}

\def\spose#1{\hbox to 0pt{#1\hss}}\def\lta{\mathrel{\spose{\lower 3pt\hbox
{$\mathchar"218$}}\raise 2.0pt\hbox{$\mathchar"13C$}}}  \def\gta{\mathrel
{\spose{\lower 3pt\hbox{$\mathchar"218$}}\raise 2.0pt\hbox{$\mathchar"13E$}}}

\begin{document}

\def\be{\begin{equation}}
\def\fe{\end{equation}}

\newcommand{\eqn}{\label}
\newcommand{\bel}{\begin{equation}\label}

\def\eqdef{\fff\ \vbox{\hbox{$_{_{\rm def}}$} \hbox{$=$} }\ \thf }

\def\ov{\overline}


\def\Lr{ {\color{rd} {L}} }
\def\Jr{ {\color{rd} {J}} }
\def\calIr{ {\color{rd} {\cal I}} }
\def\Ar{ {\color{rd} {A}} }

\def\Br{ {\color{rd} {B}} }
\def\Cr{ {\color{rd} {C}} }
\def\Dr{ {\color{rd} {D}} }

\def\Xr{ {\color{rd} {X}} }\def\Yr{ {\color{rd} {Y}} }
\def\Er{ {\color{rd} {E}} }
\def\Rr{ {\color{rd} {R}} }
\def\kappar{ {\color{rd} {\kappa}} }

\def\calMr{ {\color{rd} {\cal M}} }

\def\nablar{ {\color{rd} {\nabla}} }
\def\deltar{ {\color{rd} {\delta}} }

\def\Tr{{\color{rd} T }} 
\def\Pir{{\color{rd} {\mit\Pi} }} 
\def\calPrd{ {\color{rd} {\cal P}} }

\def\calXr{ {\color{rd} {\cal X}} }
\def\calUr{ {\color{rd} {\cal U}} }
\def\calVr{ {\color{rd} {\cal V}} }
\def\calPr{ {\color{rd} {\cal P}} }
\def\calBr{ {\color{rd} {\cal B}} }
\def\Gammar{ {\color{rd} {\Gamma}} }
\def\Psir{ {\color{rd} {\Psi}} }
\def\Sigmar{ {\color{rd} {\mit \Sigma}} }
\def\Phir{ {\color{rd} {\Phi}} }
\def\phir{ {\color{rd} {\phi}} }
\def\varphir{ {\color{rd} {\varphi}} }
\def\Thetar{ {\color{rd} {\Theta}} }
\def\thetar{ {\color{rd} {\theta}} }

\def\grd{ {\color{rd} {g}} }
\def\wrd{ {\color{rd} {\mathfrak w}} }
\def\srd{ {\color{rd} {s}} }
\def\frd{ {\color{rd} {f}} }
\def\jrd{ {\color{rd} {j}} }
\def\erd{ {\color{rd} {e}} }

\def\dr{\spose {\raise 4.0pt \hbox{\color{rd}{\,\bf-}}} {\rm d}}


\def\Euro{{\color{ru}{\spose {\lower 2.5pt\hbox{${^=}$}}{\bf C}}}}

\def\Vru{ {\color{ru} {V}} }

\def\Gru{ {\color{ru} {G}} }
\def\kru{ {\color{ru} {k}} }

\def\calAr{ {\color{ru} {\cal A}} }
\def\calGr{ {\color{ru} {\cal G}} }
\def\calCr{ {\color{ru} {\cal C}} }
\def\ConStruc{ {\color{ru} {\copyright}} }
\def\omegaru{ {\color{ru} \omega}}
\def\Omegaru{ {\color{ru} \Omega}}
\def\alpharu{ {\color{ru} \alpha}}
\def\betaru{ {\color{ru} \beta}}
\def\gammaru{ {\color{ru} \gamma}}

\def\calDr{ {\color{ru} {\cal D}} }
\def\Dru{ {\color{ru} {D}} }
\def\aru{ {\color{ru} {a}} }
\def\Aru{ {\color{ru} {A}} }
\def\Fru{ {\color{ru} F} }
\def\amr{ {\color{ru}\bf{a}} }
\def\Amr{ {\color{ru}\bf{A}} }
\def\Fmr{ {\color{ru}\bf{F}} }
\def\wrru{ {\color{ru} {\wr}} }
\def\wru{ {\color{ru} {\vert\!\!\vert\!\!\vert}} }


\def\Libra{{\color{be}{\spose {\lower 2.5pt\hbox{${^=}$}}{\cal L}}}}

\def\gbe{{\color{be} g }}
\def\kbe{{\color{be} k }}
\def\sbe{{\color{be} s }}
\def\rhob{ {\color{be} {\rho}} }
\def\varpib{ {\color{be} {\varpi}} }
\def\vb{{\color{be} v }}
\def\partialb{ {\color{be} \partial}}
\def\nablab{ {\color{be} \nabla}}
\def\Gammab{ {\color{be} \Gamma}}
\def\Deltab{ {\color{be} \Delta}}
\def\Thetab{ {\color{be} {\Theta}} }
\def\Ab{{\color{be} A }}
\def\Rb{{\color{be} R}}
\def\db{\spose {\raise 4.0pt \hbox{\color{be}{\,\bf-}}} {\rm d}}
\def\Sigmab{ {\color{be} {\mit\Sigma}} }
\def\Sb{ {\color{be} S } }

\def\calSg{\ov{\color{gn}\cal S}}
\def\calS{{\color{gn}\cal S}}
\def\ggn{{\color{gn} g}}
\def\etag{{\color{gn}\eta}}
\def\deltag{{\color{gn}\delta}}
\def\nablag{ {\color{gn} \nabla}}
\def\Kg{{\color{gn} K}} 
\def\Gammag{{\color{gn} \Gamma}} 
\def\perpg{{\color{gn}\perp\!}}
\def\xig{{\color{gn}\xi}}
\def\sigme{{\color{gy}\sigma}}

\begin{center}
{\color{rd}\bf  FIELDS IN NONAFFINE BUNDLES. III.\\[0.4cm]
Effective symmetries and conserved currents in
strings and higher branes.}
\\[1cm]
 \underline{Brandon Carter} \\[0.6cm]
 \textcolor{ru}{LUTh (CNRS),
  Observatoire Paris - Meudon. }
  \\[0.5cm]
 {\color{be} 30th November 2009.}
\\[1.2cm]
\end{center}  .

{\bf Abstract. } 
The principles of a previously developed formalism for the covariant 
treatment of multi-scalar fields for which (as in a  nonlinear 
sigma model) the relevant target space is not of affine type 
-- but curved -- are recapitulated. Their application is extended
from ordinary harmonic models to a more general category of 
{\it harmonious} field models, with emphasis on cases in which the
field is confined to a string or higher brane worldsheet, and for which 
the relevant internal symmetry group  is non Abelian, so that the 
conditions for conservation of the corresponding  charge 
currents become rather delicate, particularly when the symmetry 
is gauged. Attention is also given to the conditions for conservation 
of currents of a  different kind -- representing surface fluxes 
of generalised momentum or energy -- associated with symmetries not 
of the internal target space but of the underlying spacetime background 
structure, including the metric and any relevant gauge field. 
For the corresponding current to be conserved the latter need not be 
manifestly invariant: preservation modulo a gauge adjustment will suffice. 
The simplest case is that of ``strong'' symmetry, meaning invariance under
the action of an {\it effective} Lie derivative (an appropriately gauge 
adjusted modification of an ordinary Lie derivative). When the effective 
symmetry is of the more general ``weak'' kind, the kinetic part of the 
current is not conserved by itself but only after being supplemented by 
a suitable contribution from the background.

\vfill\eject

\section{Introduction} 
\label{Section1}

The nonlinearities most frequently encountered in classical 
field theories are broadly describable as being of three
types, of which the most common is that of coupling nonlinearity,
while the second and third types are those of kinetic nonlinearity
and target nonlinearity. Following an approach initiated in two 
preceeding articles \cite{I,II}  this article will deal with 
f nonlinearity of the {\it third} type, in which the fields under 
consideration take values in a target space that is not of the usual 
affine kind but curved.

Classical field theories of the most commonly considered kinds  
(including the familiar Yang Mills case) are kinematically linear:
the only  nonlinearity in their dynamic equations is not in the 
kinetic (meaning differential)  part, but in the purely algebraic 
 coupling contribution (which is commonly quartic in the Lagrangian 
and therefor of cubic order in the field equations).

However even if the nonlinearity of the underlying theory is only of 
this {\it first} type, various confinement mechanisms lead to 
configurations that can be treated approximately, at a less fundamental 
level by models with fewer independent degrees of freedom, but with more 
general types of non-linearity. An illustration of such a mechanism is 
provided by the prototype model set up by Witten\cite{Witten85} to 
demonstrate the possibility of conductivity in cosmic strings. This case 
furnishes an example in which the effect of confinement of the support 
zone of the field to the neighbourhood of a string or higher brane 
worldsheet is describable \cite{P92,CP95,HC08} by models of a kind 
\cite{C89,C95,LMP09} characterised by non-linearity in the gradient terms. 
Nonlinearity of this {\it second} type  has long been familiar in in 
scalar field models of the standard kind used for the treatment of
 irrotational perfect fluids and superfluids \cite{CK92,C00} as 
characterised by a generalised pressure function that plays the role of 
the Lagrangian scalar in the present work. Such kinetic nonlinearity has 
also been invoked \cite{AMS01} in some  more exotic scalar field 
theories recently  introduced in a cosmological context.

The {\it third} kind of nonlinearity arises when, instead of the support 
zone, it is the values of the field that are effectively confined
-- as for example in a reduced model \cite{C05,AMMMO06,C08,AOMM09}
due to the effect of steeply rising potential in an underlying model 
involving nonlinearity of only the first kind -- so that the result will 
be describable in the manner exemplified by non linear 
sigma models \cite{Forger92,Heus92,Heus93,Rog02,Rog04,Rog08,Ghosh06} 
in which the (differential) kinetic part enters linearly, but in which 
it is the target space of allowed field values that is nonlinear in 
the sense that the relevant structure is no longer flat but curved.

The purpose of the present work is to extend the application of 
previously developed machinery \cite{I,II} for dealing with 
nonlinearity of this {\it third} type in  multiscalar field models 
for which -- although lacking an integrable affine structure  -- the 
relevant target space, $\calXr$ say, will least at least be endowed 
with a local  affine connection. More particularly attention will be 
focussed here on the Riemannian case, for which the connection is 
derived from a metric, with components $\hat\grd_{_{\Ar\Br}}$ say, with 
respect to local coordinates $\Xr^{\!_\Ar}$ on the target space of  field 
values, so  the corresponding components of the connection will be 
{\be \hat\Gammar{^{\ _\Br}_{\!_\Ar\ _\Cr}}=\hat \grd{^{_{\Br\Dr}}}
(\hat \grd_{_\Dr(_\Ar,_\Cr)} -\frac{_1}{^2}\hat \grd_{_{\Ar\Cr},_\Dr})
\label{III.1}\fe}
using a comma to indicate partial differentiation with respect
to the coordinates, and using round brackets to denote 
index symmetrization.

In the preceeding work \cite{I,II}, the field $\Phir$ say, under
consideration was a mapping 
{\be \Phir:\ \  {\cal M} \mapsto \calXr \label{III.2} \fe}
from an n dimensional support space ${\cal M}$ endowed with its own metric
and connection, with components $\gbe_{\mu\nu}$ and
{\be \Gammab_{\!\mu\ \rho}^{\ \nu}=\gbe^{\nu\sigma}
(\gbe_{\sigma(\mu\, ,\,\rho)}-\frac{_1}{^2}\gbe_{\mu\rho\, ,\,\sigma})
\, , \label{III.3}\fe} 
with respect to local base coordinates for $x^\mu$, $\mu=0,1, ... n-1$.
The idea was that in typical applications ${\cal M}$ would represent
ordinary space time, with $n$=4, or perhaps the higher dimensional
spacetime of superstring theory, with $n$=10. 

As well as interest in models with even higher
 dimension, $n$=11, more recent developments 
have been particularly concerned with the  ubiquitous 
role of $p$-branes of various kinds, meaning subsystems confined 
to a supporting worldsheet of dimension $d=p+1$, starting with 
the case of a cosmic string, for which $p=1$.
In view of this development, the present work will be concerned
with cases in which the support of the multiscalar field 
under consideration does not extend over the whole of ${\cal M}$
but is resticted to an embedded worldsheet, $\calS$ say. 

After a recapitulation in Sections \ref{Section2}, \ref{Section3}, 
\ref{Section3a} and \ref{Section4}  of the necessary machinery
\cite{I,II,C01}, it will first be applied in Section \ref{Section5} 
to a previously considered category \cite{II} of ``forced-harmonic'' 
models that are kinetically linear, involving non linearity of the 
first type in a self coupling term, as well as non linearity of the 
third kind in a kinetic term of the harmonic kind. A category of 
``harmonious'' brane supported models  involving linearity of the 
second as well as the third (but not the first) type will then be 
introduced in Sections \ref{Section6}, and the conservation of charge 
fluxes associated with internal symmetries therein will be studied in 
\ref{Section6a}. The final sections \ref{Section7} and \ref{Section8} 
will be concerned with conservation of energy momentum fluxes 
associated with underlying spacetime background symmetries of various 
weak and strong kinds, the latter referring to invariance under the 
action of a gauge covariant modification of a Lie derivative.

\vfill\eject

\section{The bitensorial field gradient} 
\label{Section2}

To distinguish quantities pertaining to the brane worldsheet $\calS$ 
from their analogues with respect to the background ${\cal M}$ we shall 
use an overline, as in the example of the induced metric, which is
 given with respect to local brane coordinates 
$\sigme^i$ (for $ i=0, .., p-1$) by
{\be \ov\ggn_{ij}=\gbe_{\mu\nu}\, x^\mu_{\ ,i}\, x^\nu_{\ ,j}\, ,
 \label{III.4}\fe}
and which has a contravariant inverse,  $\ov\ggn{^{\,ij}}$ , whose projection
into the background provides the (first) {\it fundamental tensor} of the
imbedding, \cite{C01,Emparan09} namely
{\be \etag^{\mu\nu}=\ov\ggn{^{\, ij}}\,
x^\mu_{\ ,i}\, x^\nu_{\ ,j} \, .\label{III.5}\fe}

The preceeding work \cite{I,II} was concerned with a multicomponent
scalar field $\Phir$ defined over ${\cal M}$ so that in terms of 
 local coordinates $\Xr^{\!_\Ar}$ on the target space 
$\calXr$ its -- generically non-tensorial -- components 
$\Xr^{\!_\Ar}\{x\}$ will have tensorially transforming derivatives,
expressible as
{\be \Phir^{_\Ar}_{\ \mu}=\nablab_{\!\mu}\Xr^{\!_\Ar}\, .\label{III.6} \fe}
However such a bitensorial gradient tensor will not always be well defined
in the contexts to be considered the present work, which will be
concerned with the case of a field $\ov\Phir$ having support confined 
to a lower dimensional worldsheet $\calS$, so that it will have components  
$\Xr^{\!_\Ar}\{\sigme\}$ only for $\sigme\in\calS$.  This means
that instead of (\ref{III.6}) its gradient bitensor will have the
more restricted form
{\be \ov\Phir^{_\Ar}_{\ \mu}=\ov\nablag_{\!\mu} \Xr^{\!_\Ar}\label{III.7} \fe}
using the notation
{\be\ov\nablag_{\!\mu}=\etag_\mu^{\ \nu}\,\nablab_{\!\nu}\label{III.8}\fe}
for the relevant surface-tangential differentiation operator.
In terms of the corresponding, surface gradient operator
$\ov\nablag_{\! i}$ --  as defined in terms
of the surface coordinates $\sigme$ with respect to the induced
metric $\ov\ggn_{ij}$ -- the formula (\ref{III.7}) is equivalently
expressible in contravariant (meaning index raised) form as the 
projection
{\be  \ov\Phir^{_\Ar \, \mu}=x^\mu_{\ ,i} \, \ov\Phir^{_\Ar\, i}
\, ,\label{III.9}\fe}
where, as the worldsheet confined analogue of (\ref{III.6}),
the components
{\be \ov\Phir^{_\Ar}_{\,\ i}=\ov\nablag_{\! i}\Xr^{\!_\Ar}
\, .\label{III.10} \fe}
are bitensorial in the sense of being tensorial both with respect
to the target space coordinates $\Xr^{\!_\Ar}$ and with respect to 
the worldsheet coordinates $\sigme^i$.

\section{Gauge connection} 
\label{Section3}

If there is no symmetry group action on the target space, $\calXr$,
then it is evident that there will be no ambiguity in the 
specification of the gradient bitensors as introduced above. 
However in order to obtain a gradient operator that is well defined 
when the target space  $\calXr$ is invariant under a differential 
action, it will be necessary to specify an appropriate gauge 
connection on the corresponding fibre bundle $\calBr$, in which 
each fibre has the form of the target space $\calXr$, and in which the 
field $\Phir$ will have the status of a section over the base space 
${\cal  M}$. For this purpose -- as  discussed in more detail in the 
preceeding work \cite{I} -- the underlying background space ${\cal M}$ 
needs to be endowed, not just with its own metric $\gbe_{\mu\nu}$, but 
also with a gauge form $\Amr_\mu$ having values in the Lie algebra 
$\calAr$ of the symmetry group of the fibre space $\calXr$. 

The role of the gauge form -- as represented by vector field components
$\Aru_\mu^{\ _\Ar}$ over $\calBr$ -- is to express the deviation of 
horizontality with respect to the local fibre coordinates $\Xr^{\!_\Ar}$  
from horizontality with respect to the connection. This means that the 
effect of an infinitesimal fibre coordinate change 
$\Xr^{\!_\Ar}\mapsto \Xr^{\!_\Ar}+\delta\Xr^{\!_\Ar}$ induced by 
a fibre displacement field $\hat\kru{^{_\Ar}}= -\delta\Xr^{\!_\Ar}$
will be to map the connection form to a new value given with respect to 
the new coordinates by an affine transformation, $\Aru_\mu^{\ _\Ar}
\mapsto \Aru_\mu^{\ _\Ar}+ \delta[\hat\kru]\,  \Aru_\mu^{\ _\Ar}$, that
will be given by 
 {\be \delta[\hat\kru]\,  \Aru_\mu^{\ _\Ar}= \hat\kru{^{_\Ar}}_{,\,\mu}
-\hat\kru{^{_\Ar}}_{,_\Br} \Aru_\mu^{\ _\Br}\, .\label{III.11a}\fe} 
As the effect of the displacement on the
old connection component values will be given simply by
$\Aru_\mu^{\ _\Ar}\mapsto \Aru_\mu^{\ _\Ar}+\Aru_{\mu\ ,_\Br}^{\ _\Ar}
\,\delta\Xr^{\!_\Ar}$, it can be seen that, with respect to a fixed 
coordinate system, the net gauge change, 
 {\be \hat\deltar[\hat\kru]\,  \Aru_\mu^{\ _\Ar}=
 \delta[\hat\kru]\,  \Aru_\mu^{\ _\Ar}-\Aru_{\mu\ ,_\Br}^{\ _\Ar}
\,\delta\Xr^{\!_\Ar}\, ,\label{III.11b}\fe}
induced at a fixed position in the bundle $\calBr$ by the displacement 
$\hat\kru{^{_\Ar}}$ will be given by
{\be \hat\deltar[\hat\kru]\,  \Aru_\mu^{\ _\Ar}=\hat\kru{^{_\Ar}}_{,\,\mu}+
[ \Aru_\mu\, ,\hat\kru]^{_\Ar}\, ,\label{III.10a}\fe}
where the square bracketted term  denotes the Lie derivative of 
$\Aru_\mu^{\ _\Ar}$ with respect to $\hat\kru{^{_\Ar}}$,  namely 
{\be [\Aru_\mu\, ,\hat\kru]=- [\hat\kru\, ,\Aru_\mu]=
\Aru_{\mu\ ,_\Br}^{\ _\Ar}\hat\kru{^{_\Br}}
-\hat\kru{^{_\Ar}}_{,_\Br}\Aru_\mu^{\ _\Ar}\, .\label{III.10b}\fe}
This infinitesimal variation formula would be valid for an arbitrary
fibre tangent vector field, but for preservation of the condition
that  $\Aru_\mu^{\ _\Ar}$ should belong to the symmetry algebra
it is to be understood that $\hat\kru{^{_\Ar}}$ should also be
a symmetry generator, and therefore that it should be a solution of
the target space Killing equation
{\be \hat\nablar{^{(_\Ar}}\hat\kru{^{_\Br)}}=0\, , \label{III.81}\fe}
in which the round brackets indicate index symmetrization, where 
$\hat\nablar_{\!_\Ar}$ is the operator of covariant differentiation with 
respect to the metric $\hat\grd_{_{\Ar\Br}}$ and the
corresponding connection (\ref{III.1}) on $\calXr$.

The requirement that the gauge form  should be a target space
symmetry  generator means that its components will be expressible as
{\be \Aru_\mu^{\ _\Ar}=\Aru_\mu^{\ \alpharu}\, \aru_\alpharu^{_\Ar}
\label{III.11} \fe}
in terms of a basis $\amr_\alpharu^{_\Ar}$ of the algebra,
whose vector field components $\aru_\alpharu^{_\Ar}$ on the target
space are characterised themselves by the Killing  equation
{\be \hat\nablar{^{(_\Ar}}\aru_\alpharu^{_\Br)}=0\, .\label{III.12}\fe}
In terms of the commutators defined, according to the specification 
(\ref{III.10b}), as the Lie derivative of the first with respect to 
the second,
the corresponding structure constants 
$\ConStruc_{\alpharu\betaru}^{\,\ \ \gammaru}$
will be determined by the relations
{\be [\amr_\alpharu\, ,\amr_\betaru]=
\ConStruc_{\alpharu\betaru}^{\,\ \ \gammaru}
\amr_\gammaru \, .\label{III.14}\fe} 
 
The simplest nontrivial example is the case of a target space $\calXr$ 
that is a  2-sphere of radius $\hat\Rr$ say, for which, in terms of 
standard coordinates $\Xr^{_1}=\hat\thetar$, $\Xr^{_2}=\hat\varphir$,
the metric components will be given by the familiar prescription
$\hat \grd_{_{11}}=\hat\Rr{^2}$,  $\hat \grd_{_{12}}=0$, 
$\hat \grd_{_{22}}=\hat\Rr{^2}\, {\rm sin}^2\, \hat\thetar$. 
The Killing vectors of the associated standard basis for the
(in this case three-dimensional) symmetry algebra will have components 
$\aru_\alpharu^{_\Ar}$ that are given by
$\{ -{\rm sin}\,\hat\varphir, - {\rm cot}\,\hat\thetar\,{\rm cos}\,
\hat\varphir\}$ for $\alpharu={\bf 1} ,$   by $\{ {\rm cos}\,\hat\varphir, 
- {\rm cot}\,\hat\thetar\,{\rm sin}\,\hat\varphir\}$ for $\alpharu={\bf 2},$  
and finally by $\{0,1\}$ for $\alpharu={\bf 3} ,$ from which it can be seen 
that the corresponding structure constants will be given simply by
$\ConStruc_{\bf _{ 2\,3}}^{\bf\,\ \  _1} =\ConStruc_{\bf _{3\,1}}^{\bf\,\ \  _2}=
\ConStruc_{\bf _{ 1\,2}}^{\bf\,\ \  _3}= 1 $.
 
 Subject to the understanding that the basis should be {\it uniform} with 
respect to the chosen coordinates, in the sense that its realisation 
as a  fibre tangent vector field satisfies the condition
 {\be \aru_{\alpharu\ ,\,\mu}^{\ _\Ar}=0\, ,\label{III.85}\fe} 
 the curvature two form $\Fmr_{\mu\nu}$ of the gauge field will have 
basis components
{\be \Fru_{\!\mu\nu}^{\ \ _\Ar}=\Fru_{\!\mu\nu}^{\ \ \alpharu}
\aru_\alpharu^{_\Ar} \, \label{III.15}\fe}
that are given quite generally by by the familar formula
{\be \Fru_{\!\mu\nu}^{\ \ \alpharu}=2\partial_{[\mu} \Aru_{\nu]}^{\ \alpharu}+
\ConStruc_{\betaru\gammaru}^{\,\ \ \alpharu}\,\Aru_\mu^{\ \betaru}
\Aru_\nu^{\ \gammaru}\, .\label{III.16}\fe}
which means \cite{I} that its representation as a fibre space Killing 
vector field will be given directly by
{\be\Fru_{\!\mu\nu}^{\ \ _\Ar}= 2 \Aru_{[\nu\, \ ,\,\mu]}^{\ \ _\Ar}+ 2
\Aru_{[\nu}^{\, \ _\Br} \Aru_{\mu] \ ,_\Br}^{\, \ _\Ar}
\, .\label{III.17}\fe}
When subject to a gauge change of the form (\ref{III.10a})
this curvature form transforms according to the simple rule
{\be \hat\deltar[\hat\kru]\,\Fru_{\!\mu\nu}^{\ \ _\Ar}= 
[ \Fru_{\mu\nu}\, ,\hat\kru]^{_\Ar}\, ,\label{III.17a}\fe}
while (as a consequence of the Jacobi commutator identity) its 
antisymmetrised (exterior type) derivative will satisfy the Bianchi 
identity
{\be \Fru_{\![\mu\nu\, ,\,\rho]}^{\ \ _\Ar}+[\Aru_{[\rho}\, , \Fru_{\!\mu\nu]}]
^{_\Ar}=0\, .\label{III.17b}\fe}

As well as its induced metric $\ov\ggn_{ij}$, the brane world sheet
$\calS$ will evidently inherit a corresponding induced gauge field
with components
{\be \ov\Aru{_i^{\ _\Ar}}=\ov\Aru{_i^{\ \alpharu}}\, \aru_\alpharu^{_\Ar}
\label{III.18} \fe}
given by
{\be 
\hskip 1 cm \ov \Aru{_i^{\ \alpharu}}=  \Aru_\mu^{\ \alpharu}\, x^\mu_{\ ,\, i}
\, .\label{III.19} \fe}
The associated curvature two form on the worldsheet will have components 
{\be \ov\Fru{_{\!ij}^{\ \ \alpharu}}=2\partial_{[i} 
\ov\Aru{_{j]}^{\ \alpharu}}+
\ConStruc_{\betaru\gammaru}^{\,\ \ \alpharu}\,\ov\Aru{_i^{\ \betaru}}
\ov\Aru_j^{\ \gammaru}\, ,\label{III.20}\fe}
that are equivalently obtainable by the pullback formula
{\be \ov\Fru{_{\!ij}^{\ \ \alpharu}}= \Fru_{\!\mu\nu}^{\ \ \alpharu}\, 
x^\mu_{\ ,\, i}\, x^\nu_{\ ,\, j} \, .\label{III.21}\fe}

\section{Effective gradients in bundle.} 
\label{Section3a}

The introduction of a coordinate independent notion of horizontality
via the specification of the connection form $\Aru_\mu^{\ _\Ar}$ in the
fibre bundle $\calBr$ enable us to reduce the degree of dependence
on the fibre coordinates $\Xr^{\!_\Ar}$ that is involved in partial 
derivation with respect to the base coordinates $x^\mu$ by subtracting off
the part of the gradient that is merely attributable to the associated
gauge adjustment. We thereby obtain the corresponding
{\it effective gradient} operator, which will be denoted 
by a curly ${\calDr}$ symbol, or more compactly by a curly vertical bar
$\wrru$ before the relevant index, in the manner illustrated as follows 
in the case of an ordinary fibre tangent vector field with components
$\kru^{_\Ar}$, for which the effective gradient components
{\be \hat\kru{^{_\Ar}_{\ \wrru\, \mu}}={\calDr}_\mu \, \hat\kru{^{_\Ar}}
\label{III.21a}\fe}
will be defined by the prescription
{\be {\calDr}_\mu \,\hat \kru{^{_\Ar}}=\hat\kru{^{_\Ar}_{\, ,\,\mu}}-
 \delta[\Aru_\mu]\,\hat\kru{^{\ _\Ar}}\, .\label{III.21b}\fe} 
The gauge adjustment term here will simply be  minus the Lie derivative 
of $\kru^{_\Ar}$ with respect to the fibre tangent vector $\Aru_\mu$,
which means that, using the notation scheme introduced in (\ref{III.10a}),
it will be given by
{\be \delta[\Aru_\mu]\,\hat\kru{^{_\Ar}}=[\hat\kru\, ,\Aru_\mu]\, ,
\label{III.21c}\fe}
so that the result will be expressible according to (\ref{III.10a})
as {\be {\calDr}_\mu \,\hat \kru{^{_\Ar}}=
\delta[\hat\kru]\,  \Aru_\mu^{\ _\Ar}\, .\label{III.21d}\fe}

A noteworthy application of the forgoing formula is to the gauge 
curvature, for which the definition
{\be {\calDr}_\rho \Fru_{\!\mu\nu}^{\ \ _\Ar}=
\Fru_{\!\mu\nu\ ,\,\rho}^{\ \ _\Ar}-\delta[\Aru_\mu]\,
\Fru_{\!\mu\nu}^{\ \ _\Ar} \, ,\label{III.21e}\fe}
can be seen by (\ref{III.17a}) to give
{\be  \Fru_{\!\mu\nu\, \wrru\, \rho}^{\ \ _\Ar}= 
\Fru_{\!\mu\nu\, ,\,\rho}^{\ \ _\Ar}+[\Aru_\rho\, ,\Fru_{\!\mu\nu}]^{_\Ar}
 \, ,\label{III.21f}\fe}
from which it can be seen that the Bianchi identity
(\ref{III.17a}) will be expressible in this terminology simply  as
{\be  \Fru_{\![\mu\nu\, \wrru\, \rho]}^{\ \ _\Ar}=0\, .\label{III.21g}\fe}

A more remarkable  application of this formalism is to the case
of the gauge form itself, for which the defining prescription,
{\be {\calDr}_\mu \Aru_\nu^{\ _\Ar}=\Aru_{\nu\, , \, \mu}^{\ _\Ar}
-\delta[\Aru_\mu] \Aru_\nu^{\ _\Ar}\, ,\label{III.21h} \fe}
is to be evaluated using the formula (\ref{III.10a}), which gives
{\be \delta[\Aru_\mu]\,  \Aru_\nu^{\ _\Ar}=\Aru_\mu{^{_\Ar}}_{,\,\nu}+
[ \Aru_\nu\, ,\Aru_\mu]^{_\Ar}\, . \label{III.21i}\fe}
As the outcome, we obtain the memorable but not so well known theorem to
the effect that the  gauge curvature is simply the effective gradient 
of the gauge form, which is automatically antisymmetric:
{\be \Fru_{\!\mu\nu}^{\ \ _\Ar}=\Aru_\nu{^{_\Ar}}_{\wrru\, \mu}
=-\Aru_\mu{^{_\Ar}}_{\wrru\,\nu}\, .\label{III.21j}\fe}

Bearing in mind the convention (\ref{III.85}), it can be seen that 
the foregoing concept of effective differentiation can be taken over
directly  into terms of basis indices, so that we obtain
{\be {\calDr}_\mu\Aru_\nu^{\ \alpharu}=
\Fru_{\!\mu\nu}^{\ \ \alpharu} \label{III.21k}\fe}
and 
{\be {\calDr}_\rho\Fru_{\!\mu\nu}^{\ \ \alpharu}= \partialb_\rho
\Fru_{\!\mu\nu}^{\ \ \alpharu}+\Aru_\nu^{\ \betaru}
\ConStruc_{\betaru\gammaru}^{\,\ \ \alpharu}
\Fru_{\!\mu\nu}^{\ \ \gammaru}\, . \label{III.21l}\fe}
The Bianchi identity is thereby expressible as
{\be {\calDr}_{[\rho}\Fru_{\!\mu\nu]}^{\ \ \alpharu}=0
\, . \label{III.21m}\fe}

\section{Gauge covariant bitensorial derivatives.} 
\label{Section4}

Whereas Section \ref{Section3} and Section \ref{Section3a}
were mainly considered with fields (such as the fibre space symmetry 
generator with components $\hat\kru{^{_\Ar}}$) that were defined
throughout at least an open neigbourhood of the bundle $\calBr$, 
we shall now concentrate rather on fields (such as the basis components
 $\hat\kru{^{\alpharu}}$) that are defined just over the relevant 
base space ${\cal M}$ (or over a worldsheet $\calS$ therein).
In particular we shall be concerned with base space supported 
fields that are obtained as the restriction of bundle supported
fields to some particular bundle section as specified by
the target value of a multiscalar field mapping of the form
$\Phir:\ \{x^\mu\} \ \mapsto \{\Xr^{\!_\Ar}\}$ ( or in the
worldsheet case $\ov\Phir:\ \{\sigme^i\} \ \mapsto \{\Xr^{\!_\Ar}\}$)
for which it is necessary to distinguish the net gradient
operator, for which we shall use the symbol $\partialb$, from
the the corresponding operator of partial derivation with respect
to the bundle coordinates, for which we use a comma, in the manner
illustrated for the fibre tangent vector field $\hat\kru$ by the
relation
{\be \partialb_\mu \,\hat\kru{^{\Ar}}=\hat\kru{^{\Ar}_{\ ,\,\mu}}+
\hat\kru{^{\Ar}_{\,\ ,_\Br}} \nablab_{\!\mu} \Xr^{\!_\Br}
\, .\label{III.22a}\fe}

Proceeding in the same spirit as in the preceding section,
it is useful  -- for reducing the degree of fibre coordinate
dependence -- to replace such a base space gradient operator by 
an effective gradient operator, from which the corresponding gauge
adjustment has been subtracted off, so that it measures the deviation
from horizontality with respect to the connection.. Using the notation
{\be \Dru_\mu\Phir^{_\Ar}=\Phir^{_\Ar}_{\ \wru\,\mu} \label{III.22}\fe}
for the ensuing gauge covariant derivative of the section $\Phir$ 
itself, the definition
 {\be \Dru_\mu\Phir^{_\Ar}=\nablab_{\!\mu}\Phir^{_\Ar}
-\delta[\Aru_\mu]\Phir^{_\Ar} \label{III.22b} \fe}
in which the first term is the bitensorial quantity
 $\nablab_{\!\mu}\Phir^{_\Ar}=\partialb_\mu\Phir^{_\Ar}$
and the second term is given simply by
{\be \delta[\Aru_\mu]\,\Phir^{_\Ar}=-\Aru_\mu^{\ _\Ar}\, ,
\label{III.22c}\fe}
the effective gradient of the section is obtained, using the 
notation  (\ref{III.6}), simply as
{\be \Phir^{_\Ar}_{\ \wru\,\mu}=\Phir^{_\Ar}_{\ \mu}+\Aru_\mu^{\ _\Ar} \, .
\label{III.23}\fe}

For the analogous case of a field $\ov\Phir$ with support confined to 
the brane worldsheet $\calS$, the surface gauge form (\ref{III.18}) gives 
the correspondingly restricted covariant derivative
{\be \ov\Dru_i\ov\Phir^{_\Ar}=\ov\Phir^{_\Ar}_{\ \wru\,i} \label{III.24}\fe}
in the analogous form
{\be \ov\Phir^{_\Ar}_{\ \wru\,i}=\ov\Phir^{_\Ar}_{\ i}+\ov\Aru_i^{\ _\Ar} 
\, ,\label{III.25}\fe}
which is equivalently obtainable as the pullback 
{\be \ov\Phir^{_\Ar}_{\ \wru\,i}=x^\mu_{\  ,\, i} \ov\Dru_\mu
\ov\Phir^{_\Ar}=x^\mu_{\  ,\, i}\Phir^{_\Ar}_{\ \wru\,\mu}
\, , \label{III.26}\fe}
where 
{\be\ov \Dru_\mu\ov\Phir^{_\Ar}=\etag_\mu^{\ \nu} \Dru_\nu\ov \Phir^{_\Ar}
\, . \label{III.27}\fe}

When the concept of gauge covariant differentiation is extended
from the scalar field $\Phir$ to the vector field $\hat \kru$ on the
section, it is necessary to include an extra term to take account of the
fibre connection $\hat\Gammar$, so the ensuing 
covariant derivative takes the form
 {\be \Dru_\mu\hat\kru{^{_\Ar}}=\partialb_\mu\hat\kru{^{_\Ar}}+ \hat
\Gammar{_{\!_\Cr\ _\Br}^{\ _\Ar}}\Phir^{_\Cr}_{\ \wru\,\mu}
\hat\kru{^{_\Br}} -\delta[\Aru_\mu]\,\hat\kru{^{_\Ar}}
\, ,\label{III.28a}\fe}
 in which the relevant gauge adjustment has the simple tensorial form
{\be \delta[\Aru_\mu]\,\hat\kru{^{_\Ar}}= - \Aru_{\mu\ ,_\Br}^{\ _\Ar}
\,\hat\kru{^{_\Br}}\, .\label{III.28b}\fe}
(This formula for the adjustment of $\hat\kru$ by  $\Aru_\mu$ is
is to be contasted with the formula (\ref{III.11a}) for the 
non-tensorial but affine adjustment of $\Aru_\mu$ by $\hat\kru$.)

 In the strictly Riemannnian case (meaning absence of torsion)
to which this work and its immediate predecessor \cite{II}
is restricted, the outcome of the forgoing prescription can
be conveniently expressed in the form
 {\be \Dru_\mu\hat\kru{^{_\Ar}}=\nablab_{\!\mu}\hat\kru{^{_\Ar}}+
\kru{^{_\Br}}\hat\nablar_{\!_\Br}\Aru_\mu^{\ _\Ar}
\, ,\label{III.28c}\fe}
in which the part involving the connection has been separated out 
in the second term, while the first term is given by the ordinarily
covariant (not gauge covariant) differentiation operation
{\be \nablab_{\!\mu}\hat\kru{^{_\Ar}}=\partialb_\mu \hat\kru{^{_\Ar}}+
\Gammar_{\!\nu\ _\Br}^{\ _\Ar}\,\hat\kru{^{_\Ar}}\, ,\label{III.34a}\fe}
 with
{\be \Gammar_{\!\nu\ _\Br}^{\ _\Ar}=\Phir^{_\Cr}_{\ \nu} \,
\hat \Gammar{_{\!_\Cr\ _\Br}^{\ _\Ar}}\, .\label{III.34}\fe}

If $\hat\kru{^{_\Ar}}$ is defined not just on the section
$\Phir$ but  throughout an open neighbourhood on the bundle $\calBr$ 
-- as was supposed in Section \ref{Section4} -- then it can be seen
that the outome of the prescription (\ref{III.28a}) will also be 
expressible in terms of the effective gradient (\ref{III.21b}) as
 {\be \Dru_\mu\hat\kru{^{_\Ar}}={\calDr}_\mu \,\hat \kru{^{_\Ar}} +
\Phir^{_\Br}_{\ \wru\,\mu}\hat\nablar_{\!_\Br}\kru{^{_\Ar}}\, .
\label{III.28d}\fe} 
However that may be -- whether or not the vector $\ov\kru{^{_\Ar}}$
extends to a bundle field off the section -- the gauge covariant
derivative will always be expressible in the form
 {\be \Dru_\mu\hat\kru{^{_\Ar}}=\partialb_\mu \ov\kru{^{_\Ar}} 
+\omegaru_{\mu \ _\Br}^{\ _\Ar}\ov\kru{^{_\Br}} \fe}
using the new connector field
{\be \omegaru_{\mu \ _\Br}^{\ _\Ar}=\Phir^{_\Cr}_{\ \wru\,\mu}\, \hat
\Gammar{_{\!_\Cr\ _\Br}^{\ _\Ar}}+  \Aru_{\mu\ ,_\Br}^{\ _\Ar} 
\label{III.28}\fe}
that was introduced in the preceeding work \cite{I,II}.

This connector field is to be used for the construction \cite{I,II} of
gauge covariant bitensorial derivatives in the manner illustrated
by the case of the second gauge covariant derivative of the field 
$\Phir$, namely
{\be \Dru_\nu \Phir^{_\Ar}_{\ \wru\,\mu}=
\Phir^{_\Ar}_{\ \wru\,\mu\,\wru\,\nu }\label{III.29}\fe}
by the formula
{\be \Phir^{_\Ar}_{\ \wru\,\mu\,\wru\,\nu }= 
\Phir^{_\Ar}_{\ \wru\,\mu\, ;\,\nu} + \omegaru_{\nu \ _\Br}^{\ _\Ar}
\Phir^{_\Br}_{\ \wru\,\mu}\, , \label{III.30}\fe}
in which a semi-colon is used to indicate covariant derivation
of the ordinary kind, as given in terms of the background space connection 
by an expression of the familiar form
{\be \Phir^{_\Ar}_{\ \wru\,\mu\, ;\,\nu}=\partial_\nu\Phir^{_\Ar}_{\ \wru\,\mu}
-\Gammab_{\!\nu \ \mu}^{\ \rho}\Phir^{_\Ar}_{\ \wru\,\rho} 
\, .\label{III.31}\fe}
In this case there is no analogue of (\ref{III.28d}), because 
$\Phir^{_\Ar}_{\ \wru\,\mu}$ is well defined only on the section $\Phir$,
but, as  the analogue of (\ref{III.28c}), it is of course possible to
rewrite (\ref{III.30}) in the alternative form
{\be \Dru_\nu \Phir^{_\Ar}_{\ \wru\,\mu}=
\nablab_{\!\nu} \Phir^{_\Ar}_{\ \wru\,\mu}+\Phir^{_\Br}_{\ \wru\,\mu}
\hat\nablar_{\!_\Br} \Aru_{\nu}^{\ _\Ar} \label{III.32} \, ,\fe}
where the operation of bitensorially covariant (but not gauge
covariant) differentiation is specified as
{\be \nablab_{\!\nu} \Phir^{_\Ar}_{\ \wru\,\mu}=
 \Phir^{_\Ar}_{\ \wru\,\mu\, ;\,\nu}+\Gammar_{\!\nu\ _\Br}^{\ _\Ar}\, 
\Phir^{_\Br}_{\ \wru\,\mu}\, .\label{III.33}\fe}

When acting on the tangentially projected field
{\be \ov\Phir^{_\Ar}_{\ \wru\,\mu}=\etag_\mu^{\ \nu}
\,\Dru_\nu\Phir^{_\Ar}\label{III.35}\fe}
on a brane worldsheet $\calS$, one must take care to distinguish
its tangential derivative, with components given, using the
notation introduced in (\ref{III.10}), by
{\be  \ov\nablag_{\! \nu}\,\ov\Phir^{_\Ar}_{\ \wru\, \mu}=
\etag_\nu^{\ \rho}  \nablab_{\!\rho} \,\ov \Phir^{_\Ar}_{\ \wru\, \mu}
\, . \label{III.36}\fe}
from its tangentially projected derivative with components
{\be  \ov{\nablag_{\! \nu}\,\Phir^{_\Ar}_{\ \wru\, \mu}}
=\etag_\mu^{\ \sigma} \etag_\nu^{\ \rho}  \nablab_{\!\rho} \,
\ov \Phir^{_\Ar}_{\ \wru\, \sigma}\, . \label{III.37}\fe}
which will be the same only if the embedding is flat.
In general one must allow for the gradient of the first fundamental
tensor which will be given  by the  formula
{\be  \ov\nablag_{\! \mu} \etag_\nu^{\ \rho}= \Kg_{\mu\nu}^{\,\ \ \rho}
+\Kg_{\mu\ \nu}^{\,\ \rho}\, . \label{III.38}\fe}
in which the second fundamental tensor of the worldheet is defined 
\cite{C01,Emparan09} as
{\be \Kg_{\mu\nu}^{\,\ \ \rho}=  \etag_\nu^{\ \sigma}\, \ov\nablag_{\! \mu} 
\etag_\sigma^{\ \rho}\, . \label{III.39}\fe} 
In view of its symmetry and projection properties, namely
{\be \Kg_{[\mu\nu]}^{\ \ \ \rho}= 0\, , \hskip 1 cm 
\Kg_{\mu\nu}^{\,\ \ \sigma}\,\etag_\sigma^{\ \rho}=0 \, , \hskip 1 cm
\Kg_{\sigma\nu}^{\,\ \ \rho}\, \perpg_{\ \mu}^\sigma= 0
\, , \label{III.40}\fe} where the orthogonal projection tensor is
given by 
{\be \perpg_{\ \mu}^\nu= \gbe_{\ \mu}^\nu-\etag_{\ \mu}^\nu
\, , \label{III.41}\fe}
it can be seen that the difference between (\ref{III.36}) and 
 (\ref{III.37}) will be given by
 {\be \ov{\nablag_{\! \nu}\,\Phir^{_\Ar}_{\ \wru\, \mu}}=
 \ov\nablag_{\! \nu}\,\ov\Phir^{_\Ar}_{\ \wru\, \mu}-
\Kg_{\nu\ \mu}^{\,\ \rho}\ov\Phir^{_\Ar}_{\ \wru\, \rho}
\, . \label{III.42}\fe}
This distinction does not matter for the pullback onto the
worldsheet, which will be given by
 {\be \ov\nablag_{\! j}\,\ov\Phir^{_\Ar}_{\ \wru\, i} =
x^\mu_{\ ,\,i}x^\nu_{\ ,\,j} 
\ov\nablag_{\! \nu}\,\ov\Phir^{_\Ar}_{\ \wru\, \mu}=
x^\mu_{\ ,\,i}x^\nu_{\ ,\,j} \ov{\nablag_{\! \nu}\,
\Phir^{_\Ar}_{\ \wru\, \mu}}
\, , \label{III.43}\fe}
in agreement with what is obtained directly from the analogue of
(\ref{III.33}), namely
{\be \ov\nablag_{\!j} \ov\Phir^{_\Ar}_{\ \wru\,i}=
 \ov\Phir^{_\Ar}_{\ \wru\, i\, ;\, j}+\ov\Gammar_{\!j\ _\Br}^{\ _\Ar}\, 
\ov\Phir^{_\Br}_{\ \wru\,i}\, ,\label{III.44}\fe}
with
{\be \ov\Gammar_{\!j\ _\Br}^{\ _\Ar}=\ov\Phir^{_\Cr}_{\ j} \,
\hat \Gammar{_{\!_\Cr\ _\Br}^{\ _\Ar}}\, .\label{III.45}\fe}

The same considerations apply to the corresponding fully gauge
covariant derivative as given by
{\be \ov\Dru_\nu\ov\Phir^{_\Ar}_{\ \wru\, \mu}
=\etag_\nu^{\ \rho}\Dru_\rho\ov\Phir^{_\Ar}_{\ \wru\, \mu}
=\ov\nablag_{\! \nu}\,\ov\Phir^{_\Ar}_{\ \wru\, \mu}+
\etag_\nu^{\ \rho}\,\ov\Phir^{_\Br}_{\ \wru\,\mu}
\hat\nablar_{\!_\Br} \Aru_{\rho}^{\ _\Ar} \, ,\label{III.46}\fe}
and its tangental projection
{\be \ov{\Dru_\nu\Phir^{_\Ar}_{\ \wru\, \mu}}=\etag_\mu^{\ \sigma}
\ov\Dru_\nu\ov\Phir^{_\Ar}_{\ \wru\, \sigma}
=\ov\Dru_{\nu}\,\ov\Phir^{_\Ar}_{\ \wru\, \mu}-
\Kg_{\nu\ \mu}^{\,\ \rho}\ov\Phir^{_\Ar}_{\ \wru\, \rho}
\, , \label{III.47}\fe}
whose pullback {\be \ov\Dru_{\! j}\,\ov\Phir^{_\Ar}_{\ \wru\, i} =
x^\mu_{\ ,\,i}x^\nu_{\ ,\,j} 
\ov\Dru_{\! \nu}\,\ov\Phir^{_\Ar}_{\ \wru\, \mu}=
x^\mu_{\ ,\,i}x^\nu_{\ ,\,j} \ov{\Dru_{\! \nu}\,
\Phir^{_\Ar}_{\ \wru\, \mu}} \label{III.48}\fe}
agrees with what is obtained directly from the analogue of
(\ref{III.30}), namely 
{\be \ov\Dru_{\! j}\,\ov\Phir^{_\Ar}_{\ \wru\, i}=\partial_j
\ov\Phir^{_\Ar}_{\ \wru\, i}-\ov\Gammag{_{\! j \ i}^{\, k}}\,
\ov\Phir^{_\Ar}_{\ \wru\, k}+\ov\omegaru_{j \ _\Br}^{\ _\Ar}
\ov\Phir^{_\Br}_{\ \wru\, i}\, , \label{III.49}\fe}
in which the induced connection on the worldsheet is given by
the usual Christoffel formula,
{\be \ov\Gammag{_{\! i \ j}^{\, k}}=\ov\ggn{^{{jh}}}
(\ov \ggn_{h(i\, ,\, k)} -\frac{_1}{^2}\ov \ggn_{ik\, ,\, h})
\label{III.50}\fe}
and the worldsheet connection form for the fibre space is given by
{\be \ov\omegaru_{i \,\ _\Br}^{\ _\Ar}=x^\mu_{\ ,\,i} \,
\omegaru_{\mu \ _\Br}^{\ _\Ar}=\ov \Phir^{_\Cr}_{\ \wru\,\, i}\, \hat
\Gammar{_{\!_\Cr\ _\Br}^{\ _\Ar}}+  \ov\Aru_{i \ ,_\Br}^{\ _\Ar} \, .
\label{III.51}\fe}

\section{Conserved currents for forced-harmonic fields.} 
\label{Section5}

Before going on to the investigation of more general cases, let us 
consider the conservation of charge currents associated with 
internal symmetries in the prototype application of the foregoing 
formalism, as presented in the preceeding article \cite{II}.
That application was to a class of models that includes
the ordinary harmonic type, but that is generalised by allowance for
two kinds of force, of which the first is an internal bias
provided in the action by the inclusion of a scalar self coupling 
term, which can partially or completely break the symmetry -- 
if any -- of the target space. The other kind is an external force 
from  gauge coupling of whatever target space symmetry may 
remain unbroken. 

To be explicit, it is to be recapitulated that (in the absence of 
background weighting fields) such a  {\it biased-harmonic} system 
is characterised by an action integral of the form 
{\be \calIr=\int \! \Lr\{\Phir, \Dru\Phir\}\ \Vert \gbe\Vert^{1/2}
\, {\rm d}^n x \label{III.52}\fe}
over the base space ${\cal M}$, in which the Lagrangian scalar function
$\Lr$ is taken to be a quadratic function of the gradients of the
field section $\Phir$, with the gauge invariant form
{\be \Lr=-\frac{_1}{^2}\, \Phir_{_\Ar}{^{\wru\,\mu}}\,
\Phir^{_\Ar}{_{\wru\,\mu}} - \hat\calVr\{\Phir\}\, ,\label{III.53}\fe}
where, like the metric $\hat\grd_{_{\Ar\Br}}$ that is used for target space 
index lowering, , the potential $\hat\calVr$ is given as a fixed field on 
the space $\calXr$ in which the values of $\Phir$ are located. Nonlinear 
$\sigma$ models belong to the  special category for  which the structure 
of the target space is homogeneous, not just geometrically (as in the 
spherical example mentionned above) but also for the algebraic potential 
function $\hat\calVr$ which in that case must be just a constant that 
can be ignored as far as the field equations are concerned. Quite
generally, the field  $\hat\calVr$ must be invariant under the action
of the generators $\amr_\alpharu$ of the relevant symmetry algebra
(if any) which as well as satisfying the Killing equations
(\ref{III.12}) must also satisfy the conditions
{\be\aru_\alpharu^{_\Ar}\, \hat\calVr_{,\,_\Ar}=0 \, .\label{III.54}\fe} 

Whether or not any such symmetry algebra exists, the requirement that 
the integral $\calIr$ be unaffected by infinitesimal local variations 
of the field $\Phir$ can be seen \cite{II} to give field equations of 
the form
{\be  \Phir_{_\Ar}{^{\wru\,\mu}}{_{\wru\,\mu}}=\hat\calVr_{,_\Ar}
\, .\label{III.55}\fe}

For any symmetry algebra element ${\hat{\bf \kru}}$ say, with fibre
space Killing vector components
{\be \hat\kru{^{_\Ar}}=\hat\kru{^\alpharu}\,\aru_\alpharu^{_\Ar}
\, ,\label{III.56}\fe}
one can construct a corresponding current vector with components
{\be \Jr^\mu= \hat\kru{^{_\Ar}} \Phir_{_\Ar}^{\ \mu}=
\hat\kru{^\alpharu} \Jr_\alpharu^{\, \mu} \, ,\hskip 1 cm
 \Jr_\alpharu^{\, \mu}= \aru_\alpharu^{_\Ar}\, 
\Phir_{_\Ar}^{\ \mu}\, ,\label{III.57}\fe}
whose divergence,
{\be \Jr^\mu{_{;\,\mu}}=\Jr^\mu{_{\wru\,\mu}}\label{III.58}\fe}
can easily be evaluated using the field equations (\ref{III.54}).
Using these in conjunction with (\ref{III.54}) and the Killing 
condition (\ref{III.12}), the current divergence can be seen to 
be given by
{\be \Jr^\mu{_{;\,\mu}}=\Phir_{_\Ar}^{\ \mu}(\hat\kru{^{_\Ar}_{\ ,\,\mu}}
-\hat\kru{^{_\Ar}_{\ ,_\Br}}\,\Aru_\mu{^{_\Br}} +\hat\kru{^{_\Br}}
\Aru_{\mu\ ,_\Br}^{\ _\Ar})\, .\label{III.59}\fe}
It follows that in order for  the current to be conserved,
{\be \Jr^\mu{_{;\,\mu}}=0\, .\label{III.60}\fe}
the variation of the symmetry generator ${\hat{\bf \kru}}$
over the base space ${\cal M}$ must be restricted to satisfy
the condition
{\be \hat\kru{^{_\Ar}_{\ ,\,\mu}} =[\hat \kru, \Aru_\mu]{^{_\Ar}}
\, ,\label{III.61} \fe}
of which the right hand side is the Lie derivative
of the  symmetry generator ${\hat{\bf \kru}}$ with repect to the
gauge form $\Amr_\mu$. 
This condition can be seen to be interpretable as the obviously natural 
requirement that the Killing vector  ${\hat{\bf \kru}}$ should be 
transported horizontally with respect to the gauge connection,
or equivalently as the requirement that it should preserve the gauge
field in the sense that associated gauge adjustment 
(\ref{III.10a}) should vanish,  
{\be \delta[\hat\kru]\,  \Aru_\mu^{\ _\Ar}=0\, .\label{III.62a} \fe}

In cases for which the relevant bundle structure is that of a trivial 
direct product, for which there is a preferred gauge in which $\Amr_\mu=0$,
this horizontality requirement will be acheivable in the obvious way,
by simply taking ${\hat{\bf \kru}}$ to be {\it uniform} over ${\cal M}$, 
so that $\hat\kru{^{_\Ar}_{\ ,\,\mu}}=0$ in that gauge. However in general
the equation (\ref{III.61}) will be soluble only if the gauge field is
such as to satisfy an integrability condition which can be seen to be 
expressible in terms of the gauge curvature two-form $\Fmr_{\mu\nu}$ as
{\be [\hat \kru, \Fru_{\!\mu\nu}]{^{_\Ar}}=0
\, ,\label{III.63} \fe}
or equivalently, by (\ref{III.17a}), as
{\be \delta[\hat\kru]\,  \Fru_{\!\mu\nu}^{\ \ _\Ar}=0
\, .\label{III.64a}\fe}
What this means is that -- as could have been anticipated -- in order for
the corresponding current (\ref{III.57}) to be conserved, ${\hat{\bf \kru}}$
must generate a symmetry not just of the fibre metric $\hat\grd_{_{\Ar\Br}}$,
and of the scalar potential function $\hat\calVr$, but also of the
gauge field $\Fmr_{\mu\nu}$. 

If the symmetry group is Abelian -- as in the familiar case of ordinary
Maxwellian electromagnetism -- the requirement (\ref{III.63})
will evidently entail no further restriction, so that for any
generator  ${\hat{\bf \kru}}$ chosen uniformly over ${\cal M}$
-- meaning such that that $\hat\kru{^{_\Ar}_{\ ,\,\mu}}=0$ --
the corresponding current (\ref{III.57}) will automatically satisfy
the conservation law (\ref{III.60}), as it does even in the
non-Abelian case if the bundle $\calBr$ has a trivial direct product structure
$\calBr={\cal M}\times \calXr$ characterised by
a preferred gauge in which the connection form vanishes.

\section{Models with  harmonious  fields on branes.} 
\label{Section6}

Let us now consider cases involving a field
$\ov\Phir$ that has support restricted to a brane worldsheet $\calS$
of dimension $d=p+1$ say, so that as the analogue of (\ref{III.52})
the relevant action integral is given by an expression of the form
{\be \ov\calIr=\int_\calS  \ov\Lr\{\ov\Phir, \ov\Dru\ov\Phir\}\ \Vert
\ov\ggn\Vert^{1/2}\, {\rm d}^d \sigme \label{III.65}\, .\fe}

As well as the allowance for gauge coupling, the kind of Lagrangian
considered in the preceeding section generalised the
usual harmonic kind \cite {Misner78} by including the nonlinearity
of the first type embodied in the algebraic self coupling term
$\hat\calVr$ in (\ref{III.53}). However in the present section we
shall consider a generalisation of a different kind that will be 
referred to as {\it harmonious}, involving
non linearity of the second -- meaning kinetic -- type
as well as the nonlinearity of the third type that is embodied
in the curvature of the target space $\calXr$. Specificly we
shall use this term for cases for which the surface Lagrangian 
depends only on the target space metric $\hat\grd_{_{\Ar\Br}}$
and the symmetric target space tensor defined as
{\be\hat\wrd{^{_{\Ar\Br}}} =\etag^{\mu\nu} \hat\Phir^{_\Ar}{_{\wru\,\mu}}\,
\ov\Phir^{_\Br}{_{\wru\,\nu}}=\ov\Phir^{_\Ar}{_{\wru\,i}}\,
\ov\Phir^{_\Br\,\wru\,i}\, . \label{III.66}\fe}
 In a gauge such that the gauge form
vanishes at a particular point under consideration, this tensor
$\hat\wrd{^{_{\Ar\Br}}}$ will be identifiable simply as the induced metric
on the target space $\ov\calXr$.  In the absence of a gauge field, a 
harmonious model  will therefor be of the of the ordinarily elastic 
type in cases for which the target space is of dimension $p$,
and thus identifiable as the quotient with respect to a congruence
of timelike idealised particle worldlines on the worldsheet.
However it will not be an elastic model of the most general
kind, for which \cite{CQ72,C73} the specification of the elastic 
structure on $\ov\calXr$ would involve, not just
the metric $\hat\grd_{_{\Ar\Br}}$, but other predetermined
vectorial or tensorial fields as well.

For a model that is harmonious in the forgoing sense,
the generic variation of the Lagrangian will have 
the form
{\be \delta \ov\Lr=\frac{\partial\ov\Lr}{\partial\hat\grd_{_{\Ar\Br}}}
\,\delta\hat\grd_{_{\Ar\Br}} +\frac{\partial\ov\Lr}
{\partial\hat\wrd{^{_{\Ar\Br}}}}\,\delta\hat\wrd{^{_{\Ar\Br}}}
\, , \label{III.67}\fe}
in which,  as a Noether identity \cite{C01}, we shall have
{\be 2\,\frac{\partial\ov\Lr}{\partial\hat\grd_{_{\Ar\Br}}}
=-\kappar_{_\Cr}^{\,\ _\Ar}\,\hat\wrd^{_{\Br\Cr}}=
-\kappar_{_\Cr}^{\,\ _\Br}\,\hat\wrd{^{_{\Ar\Cr}}} \, ,\label{III.68}\fe}
 using  the notation
{\be \kappar_{_{\Ar\Br}}=\kappar_{_{\Br\Ar}}= -2\,
\frac{\partial\ov\Lr}{\partial\hat\wrd{^{_{\Ar\Br}}}}\, .\label{III.69}\fe}

The category of harmonious models defined in this way will evidently
include ordinary harmonic models \cite{Misner78}, which belong to
the special subcategory for which   $\ov\Lr$ is {\it linearly} 
dependent just on the scalar trace
{\be \hat\wrd=\hat\wrd_{_\Ar}\hat\wrd{^{_\Ar}}=\hat\grd_{_{\Ar\Br}}
\hat\wrd{^{_{\Ar\Br}}}\, . \label{III.70}\fe}
 A motivation for considering cases of more general kinds, 
starting  with that of nonlinear dependence just on $\hat\wrd$, 
 is that they can arise naturally -- for an underlying 
model with a kinetic part of the ordinary linear type -- from the effect 
of confining mechanisms of the kind commonly considered in the theory of 
topological defects. 

A prototypical example \cite{P92,CP95,HC08} is provided by the bosonic 
field model proposed by Witten\cite{Witten85}, as a simple example of the 
way currents can be confined to the worldsheets of cosmic strings. Such 
arise from spontaneous symmetry breaking by  string or higher 
$(d-1)$-brane type solutions that are longitudinally symmetric in the 
strong sense -- meaning that the relevant fields  are preserved by  
the action of the Killing vector generators $\kbe_i^{\, \mu}$ say of 
longitudinal (world sheet parallel) translations. Spontaneous symmetry 
breaking means that the solutions is not unique but belongs to a family 
of configurations mapped onto each other by the action of the relevant 
internal symmetry group $\calGr$. It is this family of configurations 
-- as labelled by the central value $\ov\Phir$ of the field $\Phir$ 
-- that forms the (typically curved) target space $\ov\calXr$ of 
the world-sheet confined effective effective model. The Lagrangian action 
for the effective model is obtained for such (current free) configurations 
simply by integrating the local action density over a transverse section 
of dimension $(n-d)$. 

The general idea is that, starting from such a family of strongly symmetric 
non-conducting configurations, a more extensive family of
 current carrying configurations will be obtainable by 
relaxing the condition that the fields be strictly invariant under the 
action of longitudinal translations but alowing them to have changes
generated by elements of the algebra $\calAr$ say of the symmetry.
The effective action for current carrying states is to be obtained
by integrating the result obtained by solving the field equations
on a particular transvers  $(n-d)$ dimensional section with values of
the gradients in the longitudinal directions orthogonal to the section
given by the action of the corresponding algebra elements as represented
by corresponding central values of the longitudinal gradients
$\Dru_i\Phir$ with values in the tangent space of $\calXr$. 

In simple cases such as that of the string configurations (with $d=2$) 
obtained \cite{P92,CP95,HC08} from the Abelian model introduced by 
Witten, the result will depend only on the trace $\hat\wrd$, albeit non 
linearly, (contrary to the oversimplified ansatz originally proposed 
by Witten himself \cite{Witten85}). The result will still depend only on 
the trace $\hat\wrd$ in the non Abelian case obtained from the minimal 
extension of the Witten model that  will be presented in a following 
article \cite{IV} in which it will be shown that this extension will give 
rise to current carrying strings supporting fields for which the  target 
space $\ov\calXr$ will have the 2-spherical form  enviseaged in Section 
\ref{Section3}.

\section{Conserved currents for harmonious fields on branes.} 
\label{Section6a}

Whatever its physical origin may be, a  Lagrangian of the harmonious
kind under consideration will have an
Eulerian (fixed point) variation that will be given,
according to (\ref{III.67}), by
{\be \delta\ov\Lr = -\frac{_1}{^2}\, \kappar_{_\Dr}^{\,\ _\Cr}\,
\wrd^{_{\Br\Dr}}\,\hat g_{_{\Br\Cr},\,_\Ar}\, \delta\Xr^{_\Ar}
-\kappar_{_{\Ar\Br}} \,\ov\Phir{^{_\Br\,\wru\,i}}
\, \delta(\Xr^{_\Ar}_{\ ,\, i}+\ov\Aru_i{^{_\Ar}}\!)
\, , \label{III.72}\fe}
which is expressible in the convenient form
{\be \delta\Lr=\frac {\delta\ov\Lr\,}{\delta\Xr^{_\Ar}}\,\delta\Xr^{_\Ar}
-\big(\kappar_{_{\Ar\Br}} \,\ov\Phir{^{_\Br\,\wru\,i}}\, 
\delta\Xr^{_\Ar}\big)_{;\, i}\, , \label{III.73}\fe}
with
{\be \frac {\delta\ov\Lr\,}{\delta\Xr^{_\Ar}}=
\big(\kappar_{_{\Ar\Br}} \,\ov\Phir{^{_\Br\,\wru\,i}}
\big)_{\wru\, i} \, .\label{III.74}\fe}

It evidently follows that, in terms of bitensorial surface current 
components defined by
{\be \ov\Jr_{\!_\Ar}{^i}=\kappar_{_{\Ar\Br}} \,\ov\Phir{^{_\Br\,\wru\,i}}
\, ,\label{III.75}\fe}
the ensuing field equations will be expressible in the neatly succinct form
{\be \ov\Jr_{\!_\Ar}{^i}_{\wru\, i}=0 . \label{III.76}\fe}
However it is to be observed that this will not in general be directly
interpretable as a current conservation law, because (unlike the last 
term in (\ref{III.73}), which is removable by integration over the base 
space) the left hand side of  (\ref{III.76})  is not an ordinary 
divergence. Nevertheless, as before, when there is an internal symmetry
group we can obtain something that actually is an ordinary divergence 
and that will vanish under appropriate conditions, by using the fact
that any fibre space symmetry generating vector field with components
$\hat\kru{^{_\Ar}}$ will define a corresponding surface charge current
with components
{\be \ov\Jr{^\nu}=x^\nu_{\ ,\, i}\,\ov\Jr{^i}\, ,\hskip 1 cm
\ov\Jr{^i}=  \hat\kru{^{_\Ar}}\, \ov\Jr_{\!_\Ar}{^i}
\label{III.77}\fe}
whose surface divergence will be given -- when the field equations are 
satisfied -- by 
{\be \ov\nablag_{\!\nu} \ov\Jr{^\nu}
=\ov\Jr{^i_{\ ;\,i}}= \ov\Jr_{\!_\Ar}{^i}
\,\hat\kru{^{_\Ar}_{\,\ \wru\,i}}\, ,\label{III.78}\fe}
where by definition we have 
{\be \hat\kru{^{_\Ar}_{\,\ \wru\,i}}= \hat\kru{^{_\Ar}_{\ ,\,i}}+ 
\hat\kru{^{_\Ar}_{\ ,\,_\Br}}\Xr^{_\Br}_{\ ,\, i}+\hat\kru{^{_\Br}}
(\Xr^{_\Cr}_{\ ,\, i}\,\hat\Gammar{_{\!_\Cr\ _\Br}^{\ _\Ar}} +
\ov\Aru{_{i\,\ ,_\Br}^{\ _\Ar}}) \, .\label{III.79}\fe}
By rewriting the latter in the form
{\be \hat\kru{^{_\Ar}_{\,\ \wru\,i}}= \hat\kru{^{_\Ar}_{\ ,\,i}}
+ \hat\kru{^{_\Br}}\, \ov\Aru{_{i\,\ ,_\Br}^{\ _\Ar}}-
\hat\kru{^{_\Ar}_{\ ,_\Br}}\ov\Aru{_i^{\ _\Br}}+
\ov\Phir{^{_\Br}_{\ \wru\,i}}\hat\nablar_{\!_\Br}\hat\kru{^{_\Ar}}
\, ,\label{III.80}\fe}
and using the symmetry property (\ref{III.68}),
one can see, as before, that provided the fibre tangent vector field
is chosen so as to satisfy the target space Killing equation
(\ref{III.81}), as well as the surface analogue of the horizontal 
transport condition (\ref{III.61}), namely
{\be \hat\kru{^{_\Ar}_{\ ,\,i}} =[\hat \kru, \ov\Aru_i]{^{_\Ar}}
\, ,\label{III.82} \fe}
we shall finally obtain a genuine surface current conservation law of the 
required kind, namely 
{\be \ov\Jr{^i_{\ ;\,i}}=0\, .\label{III.83} \fe}
In the absence of any gauge field, this can always be done for any 
element of the target space symmetry algebra. However, as before, if a
 gauge field is present, the condition (\ref{III.82}) for
 (\ref{III.83}) can only be fulfilled if the relevant integrability
condition is satisfied, namely the requirement
{\be [\hat \kru, \ov\Fru_{\!ij}]{^{_\Ar}}=0
\, ,\label{III.84} \fe}
which is interpretable as the condition that ${\hat\kru}$ should
generate a symmetry not just of the fibre metric $\hat\grd$
but also of the gauge field.

\section{Energy-momentum flux on branes.} 
\label{Section7}

The preceding section was concerned just with variations 
of the Lagrangian integral (\ref{III.65}) for {\it fixed}
values of the worldsheet location (and hence of the projection bitensor
with components $x^\mu_{\ ,\,i}$ and of the relevant background fields,
 namely the $n$-dimensional space-time metric with components 
$\gbe_{\mu\nu}$ and the gauge field as specified according to
(\ref{III.11}) in terms of components $\Aru_\mu^{\ \alpharu}$
with respect to some uniform algebra basis. Within that scheme, 
it can be seen that, when the ensuing variational field equations are
satisfied, the divergences of the corresponding basis current 
vectors with components 
{\be \ov\Jr{_{\!\alpharu}^{\, \nu}}=x^\nu_{\ ,\,i}\,\aru_{\alpharu}^{\ _\Ar}
\, \ov\Jr_{\!_\Ar}{^i}\label{III.86}\fe}
will be given , in accordance with  (\ref{III.78}), by
{\be \ov\nablag_{\!\nu} \ov\Jr{_{\!\alpharu}^{\, \nu}}= \Aru_\nu^{\ \betaru}\,
\ConStruc_{\betaru\alpharu}^{\,\ \ \gammaru}
\,\ov\Jr{_{\!\gammaru}^{\, \nu}}\, .\label{III.87}\fe}

In the present section -- generalising an approach developed for the 
Abelian case in preceeding work \cite{C95,C01} -- we shall consider the 
effect of background field variations of the form induced by 
{\it worldsheet displacements}, as generated by Lie transport with 
respect to a vector field  $\xig^\mu$, which gives 
{\be \delta \gbe_{\mu\nu}=2\nablab_{\!(\mu}\xig_{\nu)}\, ,
\label{III.88}\fe}
and 
{\be \delta\Aru_\nu^{\ \alpharu}=\xig^\mu\nablab_{\!\mu}
\Aru_\nu^{\ \alpharu}+\Aru_\mu^{\ \alpharu}\nablab_{\!\nu}
\xig^\mu\, .\label{III.89}\fe}
It can be seen that effect of this on the integrand in (\ref{III.65}) 
will be given by an expression of the form
{\be \Vert\ov\ggn\Vert^{-1/2}\delta( \ov\Lr\, \Vert\ov\ggn\Vert^{1/2})
= \frac{_1}{^2}\,\ov\Tr{^{\mu\nu}} \, \delta \gbe_{\mu\nu}
-\ov\Jr{_{\!\alpharu}^{\, \nu}}\,\delta\Aru_\nu^{\ \alpharu}\, ,
\label{III.90}\fe}
in which it can be seen that the relevant surface current coefficients will
be as given by the formula (\ref{III.86}), which can be written
more explicitely as
{\be \ov\Jr{_{\!\alpharu}^{\, \nu}}=\aru_{\alpharu}^{\ _\Ar}
\, \kappar_{_{\Ar\Br}} \,\ov\Phir{^{_\Br\,\wru\,\nu}}\label{III.91}\fe}
while the corresponding surface stress energy momentum tensor components
can be read out as
{\be\ov\Tr{^{\mu\nu}}=\kappar_{_{\Ar\Br}}\,\ov\Phir{^{_\Ar\,\wru\,\mu}}
\,\ov\Phir{^{_\Br\,\wru\,\nu}}+\ov\Lr\, \etag^{\mu\nu}\, ,
 \label{III.92}\fe}
in which it is to be recalled that $\kappar_{_{\Ar\Br}}$ will
simply be proportional to the fibre metric $\hat\grd_{_{\Ar\Br}}$
in the ordinary harmonic case, but that it will in general
depend also on  $\wrd_{_{\Ar\Br}}$, as defined by (\ref{III.66}).

Up to this point we have been treating the worldsheet location as 
something given in advance, but we shall now postulate that its
motion is governed by dynamical equations of the usual variational
type, meaning that the action (\ref{III.65}) is required to be 
preserved, not just by infinitesimal variations of the multiscalar
surface field $\ov\Phir$, but also by the arbitrary infinitesimal 
displacements generated by $\xig^\mu$. The contribution of the latter 
to the action variation can be seen -- using preceding Lie transport 
equations -- to be obtainable from  (\ref{III.90}) in the form
{\be \Vert\ov\ggn\Vert^{-1/2}\delta( \ov\Lr\, \Vert\ov\ggn\Vert^{1/2})
=\ov\nablag_{\!\nu}\Big(\xig^\mu(\ov\Tr_{\!\mu}{^\nu}\! -\Aru_\mu{^\alpharu}
\ov\Jr{_{\!\alpharu}^{\, \nu}})\Big)-\xig^\mu(\ov\nablag_{\!\nu}
\ov\Tr_{\!\mu}{^\nu}\!-\Aru_\mu{^\alpharu}\ov\nablag_{\!\nu}
\ov\Jr{_{\!\alpharu}^{\, \nu}}-2\ov\Jr{_{\!\alpharu}^{\, \nu}}\nablar_{\![\nu}
 \Aru_{\mu]}^{\,\alpharu}) \label{III.93}\fe}
of which the first part is a surface divergence that is removable
by integration. Thus when the internal field equations (\ref{III.76})
for $\ov\Phir$ are satisfied, the only remaining contribution
to the action variation will be the final, namely the contraction
with $\xig^\mu$ whose coefficient must therefor vanish.
We thus obtain a dynamical equation of the form 
{\be \ov\nablag_{\!\nu}\ov\Tr_{\!\mu}{^\nu} -\Aru_\mu{^\alpharu} \ov
\nablag_{\!\nu}\ov\Jr{_{\!\alpharu}^{\, \nu}}\!-2\ov\Jr{_{\!\alpharu}^{\, \nu}}
\nablar_{\![\nu} \Aru_{\mu]}^{\,\alpharu}=0\, .\label{III.94}\fe}
This can be conveniently rewrtten in the standard form
 {\be \ov\nablag_{\!\nu}\ov\Tr_{\!\mu}{^\nu}=\frd_\mu\label{III.95}\fe}
 in which the force density $\frd_\mu$ is a well behaved (algebra basis 
independent) covector that can be seen from (\ref{III.87}) to be 
expressible, using the definition (\ref{III.16}), in the form
{\be\frd_\mu=\ov\Jr{_{\!\alpharu}^{\, \nu}}\, \Fru_{\!\nu\mu}^{\ \ \alpharu}
= \ov\Jr{_{\!_\Ar}^{\ \nu}}\,\Fru_{\!\nu\mu}^{\, \ \ _\Ar} \, .
\label{III.96}\fe}

This expression  generalises the formula that is already familiar in 
the ordinary electromagnetic case, for which the gauge algebra is that 
of a U(1) action on the unit circle. Subject to the usual
understanding that the latter is parametrised by the angle coordinate
$\Xr^{_1}=\varphir$, our previous treatment of this Maxwellian case
\cite{C95,C01} can be expressed in terms of the formalism used 
here by setting $\Aru_\mu^{\ _1}=-\erd\Aru_\mu$ so that 
$\Fru_{\!\mu\nu}^{\ \ _1}=-\erd\Fru_{\!\mu\nu}$ and $\ov\jrd{^\mu}
=-\erd\ov\Jr{_{\!_1}^{\, \nu}}$, where $\erd$ is the relevant 
charge coupling constant. (In typical applications using unrationalised 
Planck units, the latter will be taken to be given approximately by 
$\erd=1/\sqrt{137}$, while the presence of the negative sign is 
attributable to the unfortunate but historically entrenched convention 
that for ordinary electrons the electromagnetic current direction is 
{\it opposite} to that of the particles themselves).

As in the familiar Abelian case \cite{C95,C01}, it is to be noticed that
 the tangentially projected part of the force equation (\ref{III.95}) 
provides no new information, being merely an automatic consequence 
of the internal field equations (\ref{III.76}) on the worldsheet, whereas 
the orthogonally projected part provides the extra information needed 
to determine the evolution of the string world sheet, whose equation 
of motion is thereby obtained in the standard form
{\be\ov\Tr{^{\nu\rho}} \Kg_{\nu\rho}^{\,\ \ \mu} =
\perpg^{\mu\nu}\frd_\nu \, .\label{III.97}\fe}

As was done for the charge currents considered in the preceeding section,
we can again specify a current that may be conserved by contracting the 
relevant free index with symmetry generating vector field, but this time
not on the target space but on the base ${\cal M}$, where the relevant 
Killing equation for preservation of the metric $\gbe_{\mu\nu}$ by the
vector field $\kbe^\mu$ in question takes the form
{\be \nablab^{[\mu}\kbe^{\nu]}=0\, .\label{III.98}\fe}
The corresponding current,
{\be \ov\Pir{^\mu}=\kbe^\nu\,\ov\Tr_{\!\nu}{^{\mu}}\, ,\label{III.99}\fe}
will be interpretable as a flux of  momentum  when  $\kbe^\mu$
is the generator of a spacelike translation, while  corresponding to
a flux of energy in the timelike case for which (with the sign 
convention used here) $\kbe^\mu\kbe_\mu$ is negative.It evidently 
follows from (\ref{III.95}) that its surface divergence will be given by
{\be \ov\nablag_{\!\nu} \ov\Pir{^\nu}=\kbe^\mu\frd_\mu\, ,
\label{III.100}\fe}
and thus that it will be conserved,
{\be \ov\nablag_{\!\nu} \ov\Pir{^\nu}=0\, ,\label{III.101}\fe}
when the force does no work, which by (\ref{III.96}) will be the case
if and only if the gauge field is such that
{\be \kbe^\mu\,\Fru_{\!\mu\nu}^{\, \ \ \alpharu}\,\ov
\Jr{_{\!\alpharu}^{\ \nu}}=0
\, .\label{III.102}\fe}
It is to be remarked that this requirement will always be satified
if the current (and hence also the worldsheet in which it is contained)
happens to be entirely aligned with the Killing vector, 
{\be \kbe^{[\mu}\ov\Jr{_{\!\alpharu}^{\ \nu]}}=0\, ,\label{III.103}\fe}
a condition that is describable as {\it staticity} in the case for 
which the Killing vector is timelike so that the ensuing conservation 
law is that of an energy flux. It is evident that the requirement 
(\ref{III.102}) will also hold if, instead of the current, it is the
gauge field itself that has the property describable, if the 
Killing vector is timelike, as staticity, meaning vanishing 
of its ``electric'' (as opposed to ``magnetic'') part, namely 
{\be \kbe^\mu\,\Fru_{\!\mu\nu}^{\, \ \ \alpharu}=0\, .\label{III.104}\fe}

\section{Weak, effective, strict, and strong symmetries.} 
\label{Section8}

A field over the base space ${\cal M}$ is describable as 
{\it manifestly symmetric} \cite{RaduVolkov08} with respect to 
the continuous transformation group generated by a vector field with 
components $\kbe^\mu$ on ${\cal M}$ if is invariant under the 
corresponding Lie transport operation, that is to say if it is 
mapped to zero by  the corresponding Lie differentiation operator 
$\Libra[\kbe]$, which will be given for the section $\Phir$ simply by
{\be \Libra[\kbe]\Phir^{_\Ar}=\kbe^\mu\Phir^{_\Ar}_{\,\mu}
\, .\label{III.120}\fe}
For the relevant independent background fields,namely the metric and the
basis components of the gauge field, it will be given  by
{\be \Libra[\kbe]\gbe_{\mu\nu}= 2\nablab_{\![\mu}\kbe_{\nu]} \, ,
\label{III.121} \fe}
and 
{\be \Libra[\kbe]\Aru_\mu^{\ \alpha}=\kbe^\nu\Aru_{\mu\ ,\,\nu}^{\ \alpha} 
+\kbe^\nu_{\ ,\,\mu}\Aru_\nu^{\ \alpha}\, ,\label{III.122}\fe}
while for the basis components of the gauge curvature it will be given by
{\be \Libra[\kbe]\Fru_{\mu\nu}^{\ \ \alpha}
=\kbe^\rho\Fru_{\mu\nu\ ,\,\rho}^{\ \  \alpha} +2 \kbe^\rho_{\ ,\,[\nu}
\Fru_{\mu]\rho}^{\ \ \ \alpha}\, .\label{III.123}\fe}

The apparent variation measured in this way is however highly gauge 
dependent. A more meaningful  measure of actual physical variation is 
obtainable -- as for the bitensorially gauge covariant covariant 
differentation procedure described above -- by subtracting off the 
relevant gauge adjustment as generated by the corresponding fibre 
tangent field, with components $\hat\kru_\mu{^{_\Ar}}=
\hat\kru_\mu{^{\alpharu}}\, \aru_\alpharu^{_\Ar}$ given by
{\be   \hat\kru{^{\alpharu}}
=\kbe^\mu\Aru_\mu^{\alpharu}\, .\label{III.124}\fe}
This provides what we shall refer to as the {\it effective Lie  
derivative}, which we shall distinguish from its ordinary analogue by 
use of the financial euro symbol in place of the traditional libra 
symbol according to the prescription
{\be \Euro[\kbe]=\Libra[\kbe]-\delta[\hat\kru]\, .\label{III.125}\fe}
The required gauge adjustments will be given for the section
and the metric simply by
 {\be \delta[\hat\kbe]\Phir^{_\Ar}=-\hat\kru^{_\Ar}\, ,\hskip 1 cm
\delta[\hat\kbe]\gbe_{\mu\nu}=0\, ,\label{III.126}\fe}
so for the latter there is no difference between ordinary and
effective Lie differentation while for the section, as the 
analogue of (\ref{III.120}), in the notation of  (\ref{III.22})
we simply get 
{\be \Euro[\kbe]\Phir^{_\Ar}=\kbe^\mu\,\Phir^{_\Ar}_{\ \wru\,\mu} 
\, ,\label{III.127}\fe}
For the gauge field we have the less trivial adjustment
 {\be \delta[\hat\kru]\Aru_\mu^{\ \alpharu}=\hat\kru{^{\alpharu}_{\, ,\,\mu}}
+\Aru_\mu^{\ \betaru}\ConStruc_{\betaru\gammaru}^{\,\ \ \alpharu}
\,\hat\kru{^{\gammaru}} \, ,\label{III.128}\fe}
which leads however to the neat and memorable result
{\be \Euro[\kbe]\Aru_\mu^{\ \alpharu}=\kbe^\nu
{\calDr}_\nu\Aru_\mu^{\ \alpharu}=\kbe^\nu\Fru_{\nu\mu}^{\ \ \alpharu}
\, ,\label{III.129}\fe}
while for the gauge curvature we have
{\be \delta[\hat\kru]\Fru_{\mu\nu}^{\ \ \alpharu}=
\Fru_{\!\mu\nu}^{\ \ \betaru}\ConStruc_{\betaru\gammaru}^{\,\ \ \alpharu}
\,\hat\kru{^{\gammaru}} \, ,\label{III.128a}\fe}
which leads, via the Bianchi identity (\ref{III.21m}), to
{\be \Euro[\kbe]\Fru_{\mu\nu}^{\ \ \alpharu}=
2{\calDr}_{[\nu}(\kbe^\rho\Fru_{\mu]\rho})\, .\label{III.129a}\fe}

Just as a field configuration may be said to be manifestly symmetric, 
with respect to a displacement generator $\kbe^\mu$, if the 
corresponding Lie derivative vanishes, the configuration will 
be similarly describable as {\it strongly symmetric} with respect to 
$\kbe^\mu$ if the corresponding {\it effective} Lie derivative is
zero. However it will be describable as merely {\it weakly symmetric}
if this effective Lie derivative does not vanishes absolutely, 
but only modulo the  action of some internal symmetry generator with 
base components $\Vru^\alpharu$ say, or equivalently if the 
{\it ordinary} Lie derivative vanishes modulo the action of the 
difference $\Vru^\alpharu-\hat\kru{^\alpharu}$, with
$\hat\kru{^\alpharu}$ as defined by (\ref{III.123}).
It is to be remarked that manifest symmetry need only be of the
weak kind when a non-integrable gauge field is present, but that
it will be of the strong kind when such a field is absent.

When applied to something as simple as a scalar section $\Phir$,
the  weak symmetry condition, 
{\be \Euro[\kbe]\Phir^{_\Ar}+\delta[\Vru]\Phir^{_\Ar}=0
\, , \label{III.130b}\fe}
can be seen, from the formula $\delta[\Vru]\Phir^{_\Ar}
=-\Vru^{\!_\Ar}$, to  reduce to an equation of the form
{\be \kbe^\nu \Phir^{_\Ar}_{\ \wru\,\nu}=\Vru^{\!_\Ar}
\, .\label{III.130c}\fe}
However this will entail no restriction at all if the symmetry 
group is transitive over the target space (as for example when the 
latter is spherically symmetric) as it will be trivially soluble 
for $\Vru^{\!_\Ar}$ as a space-time position
dependent target-space Killing vector on the section. 

A more meaningful condition that may appropriately be described as
{\it strict} symmetry is that of a weak symmetry for which the
relevant adjustment is restricted to be such as to preserve 
the connection. In other words a configuration will be describable
as {\it strictly symmetric} with respect to 
$\kbe^\mu$ if the effect on it of the corresponding {\it effective} 
Lie derivative can be cancelled by the action of some
internal symmetry generator with base components $\Vru^\alpharu$
such that $\delta[\Vru]\Aru_\mu^{\ \alpharu}$ vanishes, which,
according to (\ref{III.10a}),  means
that it must satisfy the horizontal transport equation
{\be \partialb_\mu \Vru^\alpharu+\Aru_\mu^{\ \betaru}
\ConStruc_{\betaru\gammaru}^{\,\ \ \alpharu}\Vru{^\gammaru}
=0\, ,\label{III.130d}\fe}
which, as discussed in Section \ref{Section5}, will be integrable 
only if the curvature satisfies the corresponding condition
{\be \Fru_{\!\mu\nu}^{\,\ \betaru}
\ConStruc_{\betaru\gammaru}^{\,\ \ \alpharu}\Vru{^\gammaru}
=0\, .\label{III.130e}\fe}

A less restrictive but still meaningful condition that may be described
as {\it effective symmetry} is obtained by relaxing the foregoing
condition of horizontal transport in all directions to that
of horizontal transport just in the direction of the Killing vector.
In other words a configuration will be describable
as {\it effectively symmetric} with respect to 
$\kbe^\mu$ if the effect on it of the corresponding {\it effective} 
Lie derivative can be cancelled by the action of some
internal symmetry generator that is itself {\it strongly} symmetric, 
meaning that its base components $\Vru^\alpharu$ satisy the requirement
 {\be \Euro[\kbe]\Vru{^\alpharu}=0 \fe}
in which it is to be recalled that, by definition, we shall have
 {\be \Euro[\kbe]\Vru{^\alpharu}= \kbe^\nu\, \delta[\Vru]\Aru_\nu^
{\ \alpharu}= \kbe^\nu(\partialb_\nu \Vru^\alpharu+\Aru_\mu^
{\ \betaru}\ConStruc_{\betaru\gammaru}^{\,\ \ \alpharu}\Vru{^\gammaru}
)\, .\label{III.130f}\fe}
In the particular case of the section $\Phir$ it is to be remarked
that effective symmetry in the foregoing sense is equivalent to the 
postulate of strong symmetry of its gauge covariant derivative 
$\Phir^{_\Ar}_{\ \wru\,\nu}$.

Various kinds of symmetry in the categories defined above were studied 
in work by Forgacs and Manton \cite{ForMan80}, albeit with limited 
generality, in that these authors considered only   target space 
symmetries that were ``gauged'' in the sense that the physical presence 
of a nonintegrable connection field was admitted  by the theoretical 
model under consideration, whereas for strict symmetry of the most 
general kind \cite{RaduVolkov08} a target space symmetry that is not 
in the gauged subalgebra but merely ``global'' will  also be perfectly 
acceptable.

The most important application of these successively more restrictive
notions of weak, effective, strict, and strong symmetry is of course
to the gauge field itself. In this particular case the distinction 
between strict and strong symmetry disappears, as the former condition
will automatically entail the latter, namely
{\be \Euro[\kbe]\Aru_\nu^{\ \alpharu}=0 \, .\fe}
It can be seen from (\ref{III.129}) that this strong symmetry condition
is equivalent to the sufficient condition (\ref{III.104}) for the 
generalised surface momentum flux conservation property (\ref{III.101}).
This sufficient condition for conservation of the current characterised 
by $\kbe^\mu$ is thus interpretable as the requirement that, as well as 
satisfying the Killing equation (\ref{III.98}), this vector field  
should generate a strong symmetry of the gauge field.

In the case of the gauge field (unlike that of the section $\Phir$)
 symmetry of even  the weak type has non trivial consequences.
The meaning of weak symmetry for the gauge field is the possibility of 
constructing what is describable as a generalised voltage field, 
consisting of some fibre space symmetry generator, with basis 
components $\Vru{^\alpharu}$ such that
 {\be  \Euro[\kbe]\Aru_\nu^{\ \alpharu}+
\delta[\Vru]\Aru_\nu^{\ \alpharu}=0 \, .\label{III.130}\fe}
As a necessary integrability condition for this, it can be seen that 
a weak symmetry condition of the same form with the same voltage field 
$\Vru{^\alpharu}$ must also be satisfied by the gauge curvature, for 
which we thus obtain the requirement
{\be \Euro[\kbe]\Fru_{\mu\nu}^{\ \ \alpharu}=
\Vru{^{\betaru}}\ConStruc_{\betaru\gammaru}^{\,\ \ \alpharu}
\Fru_{\!\mu\nu}^{\ \ \gammaru}\, \, .\label{III.130a}\fe}

The weak  symmetry condition (\ref{III.130}) can evidently be 
rewritten in the form
 {\be  \Libra[\kbe]\Aru_\nu^{\ \alpharu}=
\delta[\hat \kru\!-\!\!\Vru]\Aru_\nu^{\ \alpharu} \label{III.131}\fe}
which makes it apparent how, as remarked above, manifest symmetry is 
interpretable  as the special case of weak symmetry for which 
$\Vru{^\alpharu}$ is equal  to $\hat\kru{^\alpharu}$ as given by 
(\ref{III.124}), whereas strong symmetry is the special case for which 
the relevant voltage field $\Vru{^\alpharu}$ simply vanishes.

By writing out the condition (\ref{III.130}) of weak symmetry of the 
gauge field in the explicit form
{\be \kbe^\mu\Fru_{\!\mu\nu}^{\ \ \alpharu}+\Vru{^\alpharu_{\! ,\,\nu}}+
\Aru_\nu^{\ \betaru}\ConStruc_{\betaru\gammaru}^{\,\ \ \alpharu}
\Vru{^\gammaru} =0 \, ,\label{III.132}\fe}
it can be seen to imply that the surface current divergence condition
(\ref{III.100}) will be expressible as
{\be \ov\nablag_{\!\nu} \ov\Pir{^\nu}=-\big(
\Vru{^\alpharu_{\! ,\,\nu}}+\Aru_\nu^{\ \betaru}
\ConStruc_{\betaru\gammaru}^{\,\ \ \alpharu}\Vru{^\gammaru}\big)
\ov\Jr{_{\!\alpharu}^{\ \nu}}\, .\label{III.133}\fe}
Under these circumstances it can be seen from the generally valid 
current divergence formula (\ref{III.87})
that we shall obtain a strict surface current conservation law, of the form
{\be \ov\nablag_{\!\nu} \ov{\calPrd}{^\nu}=0\, ,\label{III.134}\fe}
by setting
{\be \ov{\calPrd}{^\nu}=-\ov\Pir{^\nu}+
\Vru{^\alpharu}\ov\Jr{_{\!\alpharu}^{\, \nu}} \, ,\label{III.135}\fe}
in which  both $\ov{\Pir}{^\nu}$
and $\Vru{^\alpharu}$ depend on the choice of the Killing field 
$\kbe^\nu$. In the case for which this Killing vector is a time
translation generator, so that the contribution $-\ov\Pir{^\nu}$ will 
be interpretable as a flux of kinetic energy, the extra term 
$\Vru{^\alpharu}\ov\Jr{_{\!\alpharu}^{\, \nu}}$ in (\ref{III.135}) will 
be interpretable as a corresponding flux of potential energy, while 
the voltage field $\Vru{^\alpharu}$ can be seen to be the natural 
non-Abelian generalisation of an ordinary electrostatic potential 
field in Maxwellian electromagnetism. In the special case for which the 
section itself satisfies the weak symmetry condition  (\ref{III.130c}),  
this conserved total energy flux will simply be
$\ov{\calPrd}{^\nu}=-\ov\Lr \, \kbe^\nu$, and if the symmetry 
thus generated by  $\kbe^\nu$ is not merely weak but {\it strict} 
the right hand side of (\ref{III.133}) will vanish, which means that 
both the kinetic contribution  $-\ov\Pir{^\nu}$ and the potential 
contribution $\Vru{^\alpharu}\ov\Jr{_{\!\alpharu}^{\, \nu}}$ will be 
separately conserved.

\bigskip
{\bf Acknowledgements}
\medskip

 The author is grateful to Marc Lilley, Xavier Martin, Patrick
 Peter, and  Mikhail Volkov for stimulating conversations.

\end{document}